\documentclass[aps,prb,twocolumn, superscriptaddress]{revtex4-2}

\usepackage[dvipsnames]{xcolor}
\definecolor{darkblue}{rgb}{0,0,.65}

\usepackage{microtype}
\usepackage{graphicx}
\usepackage[export]{adjustbox}
\usepackage{pifont}

\usepackage[
    pdfauthor   = {Ophelia Sommer},
    pdfstartview= FitH,
    breaklinks  = true,
    bookmarks   = true,
    colorlinks  = true,
    anchorcolor = black,
    citecolor   = blue,
    filecolor   = red,
    menucolor   = black,
    urlcolor    = blue,
    linkcolor   = blue,
]{hyperref}
\usepackage[all]{hypcap}

\usepackage{natbib}

\usepackage{amsmath,amssymb,amsthm}
\usepackage{siunitx}
\usepackage{braket} 
\usepackage{bm}
\usepackage{mathrsfs}
\usepackage{booktabs}

\makeatletter
\newcommand\notsotiny{\@setfontsize\notsotiny{6.5}{7.6}}

\usepackage{tikz}
\usepackage[normalem]{ulem}
\usetikzlibrary{decorations.markings}
\usetikzlibrary{calc,backgrounds,decorations.pathmorphing,decorations.pathreplacing}
\pgfdeclarelayer{background}
\pgfdeclarelayer{foreground}
\pgfdeclarelayer{super-foreground}
\usetikzlibrary{arrows}

\pgfsetlayers{background,main,foreground,super-foreground}

\tikzstyle{tensor}=[rectangle,rounded corners=2,draw=black]
\tikzstyle{tensorcirc}=[circle,draw=black]
\let\oldi\i
\newcommand{\Z}{\mathbb{Z}}
\newcommand{\R}{\mathbb{R}}

\newcommand{\e}{\mathrm{e}}
\renewcommand{\i}{\mathrm{i}}
\newcommand{\dd}{\mathrm{d}}

\newcommand{\mc}[1]{\mathcal{#1}}
\newcommand{\mr}[1]{\mathrm{#1}}
\newcommand{\bb}[1]{\mathbb{#1}}

\DeclareMathOperator{\tr}{Tr}
\let\Im\relax
\DeclareMathOperator{\Im}{Im}

\newcommand{\vb}[1]{\mathbf{#1}}

\theoremstyle{definition}

\def\be{\begin{equation}}
\def\ee{\end{equation}}

\begin{document}
\title{Higher Berry Curvature from the Wave Function I: \\ Schmidt Decomposition and Matrix Product States}

\author{Ophelia Evelyn Sommer}
\affiliation{Department of Physics, Harvard University, Cambridge MA 02138}

\author{Xueda Wen}
\affiliation{School of Physics, Georgia Institute of Technology, Atlanta, GA 30332, USA}
\affiliation{Department of Physics, Harvard University, Cambridge MA 02138}
\affiliation{Department of Physics, University of Colorado, Boulder, CO 80309, USA}

\author{Ashvin Vishwanath}
\affiliation{Department of Physics, Harvard University, Cambridge MA 02138}

\date{\today}
\begin{abstract}
	Higher Berry curvature (HBC) is the proposed generalization
	of Berry curvature to infinitely extended systems. Heuristically HBC
	captures the flow of local Berry curvature in a system.
	Here we provide a simple formula for computing the HBC for extended $d=1$ systems
	at the level of wave functions using the Schmidt decomposition.
	We also find a corresponding formula for matrix product states (MPS),
	and show that for translationally invariant MPS this gives rise to a quantized invariant.
	We demonstrate our approach with an exactly solvable model and numerical calculations
	for generic models using iDMRG.
\end{abstract}

\maketitle
\section{Introduction}
The central object in the study of the geometry of quantum states is the Berry curvature\cite{berry84}.
The integral of this curvature over a closed surface is the quantized invariant
that gives rise to many
topological phenomena such as the integer quantum hall effect\cite{thoulessQuantizedHallConductance1982a}.
In order to extend the notion of quantum geometry to (gapped) many body systems, it is imperative to find a generalization of this curvature, which similarly has a quantized integral, even in infinitely large systems. A naive generalization, would be the total Berry curvature of the entire system, but this curvature is extensive, and so the total is ill defined. Worse still, in infinite systems, even in the case this total is finite, we can change the total Chern number of a system, with finite time evolution.
A similar problem arises in the more familiar case of a system with a  global $\mr{U}(1)$ symmetry. Indeed, the total charge of a finite system gives an invariant. In an infinite system, the total charge is extensive, and so need not be finite. Further, finite time evolution can create a flow of charge out to one of the edges at infinity. In the $\mr{U}(1)$ case this flow out to infinity can be a blessing, since it gives rise to a new topological invariant over a family of states parameterized by a circle $S^1$ in one dimension; the well known Thouless pump\cite{thouless_1983}. Initiated by Kitaev\cite{kitaev2019}, an analogous development has occurred with Berry curvature\cite{ KS2020_higherberry, KS2020_higherthouless, Hsin_2020, Cordova_2020_i, Cordova_2020_ii, Else_2021,Choi_Ohmori_2022, Aasen_2022, qpump2022,
	Hsin_2023,Kapustin2201, Shiozaki_2022, 2022aBachmann,Ohyama_2022, ohyama2023discrete, Kapustin2305, homotopical2023, 2023Ryu, 2023Qi, 2023Shiozaki,2023Spodyneiko,2023Debray}.
This higher Berry curvature\cite{KS2020_higherberry} quantifies the flow of local Berry curvature around the many body system,
which leads, for instance, to Chern number pumping\cite{qpump2022}.
A computation of this higher Berry curvature, which is a $d+2$ form in $d$ dimensions\footnote{This is analogous to how a
	$p$ form symmetry current in spatial dimension $d$ will be a
	closed $d-p$ form. The constructions considered here are thus to
	view the Berry curvature as a $p=-2$ form ``symmetry'' current
	existing on the joined spacetime and parameter space. The higher
	Berry invariant will be associated by integrating this current
	over the parameter space above a fixed spatial point.}, was given in \cite{KS2020_higherberry} in terms of the parent many body Hamiltonian to a ground state, and its relation to the local algebra of observables\cite{Kapustin2201}. It is expected that this curvature has quantized integrals for short range entangled states, which has been proved in $d=1$\cite{Kapustin2305}. Computing the curvature from the parent Hamiltonian requires computing expectation values of the resolvant and the local perturbations to the Hamiltonian. A natural question is thus: can the higher Berry curvature be defined {\em directly} on a family of invertible wave functions, rather than in terms of  parent Hamiltonians?
Here we show how to define the higher Berry curvature directly on states in one dimension, using their matrix product state (MPS) parameterization, and the related Schmidt decomposition. In \cite{2024long} we answer the question more generally for locally parameterized wave functions beyond $d=1$.
In particular, expressing the wave function in terms of the Schmidt
decomposition between regions $A$ and $B$,
$\ket{\psi}=\sum_{\alpha}c_\alpha\ket{\alpha}_{[A]}\ket{\alpha}_{[B]}$
the higher Berry curvature takes the elementary form
\begin{equation}
	\Omega^{(3)}=\Im\sum_{\alpha}\dd c^2_\alpha\braket{\dd \alpha|\dd\alpha}_{[A]},
	\label{eq:schmidt_hbc_intro}
\end{equation}
where given a family of states with parameters $\lambda^l$, the exterior derivative
operator takes the form
$\dd=\sum_{l}\frac{\partial}{\partial\lambda^l}\dd\lambda^l$,
and we implicitly antisymmetrize over all appearances of
$\dd\lambda^l$ as usual. This expression can be efficiently computed using MPS, and we prove its integral is quantized for uniform MPS, using the recently discovered gerbe structure of uMPS\cite{2023Ryu,2023Qi}. This provides the relation between this mathematical structure and the Berry curvature flow perspective of the invariant. Previous work\cite{2023Shiozaki} used a discretized parameter space to also compute the topological invariant, although the precise relation between their work and the HBC defined in this paper is unclear.
We note in passing a similar expression can be found for the  Thouless pump in $d=1$. In particular there is a $1$-form on parameter space, which when integrated gives the Thouless pump invariant. Using the charge operator $Q_{[A]}$ of region $A$, this form can be written $\sum_{\alpha}\dd c_\alpha^2 \braket{\alpha|Q_{[A]}|\alpha}_{[A]}$.

\section{Flows of Berry curvature}
\label{sec:MPS-constructions}
\begin{figure}
	\centering
	\includegraphics{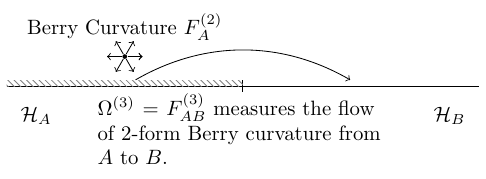}
	\caption{Higher Berry curvature in $d=1$ is the flow of regular Berry curvature
		across a partition of the system.
	}
	\label{fig:1d_HBC}
\end{figure}
Higher Berry curvature in $d=1$ spatial dimension is characterised by the flow of
the Berry curvature between regions that have boundaries at infinity\cite{qpump2022}. In particular if we consider the regular lattice $\Z$ and an arbitrary cut $a$ it will be a flow between the regions $A=\Z_{< a}$ and $B=\Z_{\geq a}$ which have `boundaries' at $\pm\infty$.
Our central question is how to assign real space locality to the Berry curvature,
and computing from this the corresponding flow.
To build up to the $d=1$ system of an infinite length, consider a finite subset of the one-dimensional lattice,
consisting of $N$ sites, such that the total Hilbert space $\mc{H}=\otimes_{p=1}^N \mc{H}_p$, with a basis $\ket{s_p}$ of $\mc{H}_p$.
We will work with a parameterized family of normalised gapped wave functions $\ket{\psi}$ living in $\mc{H}$.
We work in local neighbourhoods of parameter space, so that we can pick a smooth gauge for the wave functions, and so consider the family to be a function $X\to \mc{H}$, with
associated exterior derivatives $\dd$.
The total Berry curvature is the differential 2-form
$\Omega^{(2)}=-\Im\braket{\dd\psi|\dd\psi}=\dd \mc{A}$, where
$\mc{A}=-\Im\braket{\psi|\dd \psi}$ is the connection 1-form.
The higher Berry curvature we construct will be independent of the choice of gauge, though our intermediate expression will not be.

Suppose briefly that the family of wave functions $\ket{\psi}$ is completely unentangled
$\ket{\psi}=\ket{\psi}_{[1]}\otimes\cdots\otimes\ket{\psi}_{[N]}$.
Then there is a natural notion of the Berry curvature at site $p$, namely the Berry curvature
of the site $p$ wave function, which we denote $F_{p}^{(2)}=-\Im\braket{\dd \psi|\dd \psi}_{[p]}$.
Equivalently, this expression arises from decomposing the global variation $\dd$ into local variations
\begin{equation*}
	\dd\ket{\psi}=\sum_p \ket{\psi}_{[1]}\otimes \cdots\otimes \dd\ket{\psi}_{[p]}\otimes\cdots\otimes \ket{\psi}_{[N]}=\sum_p\dd_p\ket{\psi},
\end{equation*}
or simply written $\dd=\sum_p\dd_p$. Then using these local variation $F_p^{(2)}=\dd_p\mc{A}$.
To generalize this perspective to the short-range entangled case, we must define what the derivative operator $\dd_p$ means. In particular, the decomposition will need to be local, so that (1) $\dd_p \mc{A}=-\Im\braket{\dd\psi|\dd\psi}_{[p]}$ in the unentangled case, (2) in the entangled case we require that $\dd_p\dd_q\mc{A}$ decays exponentially in the distance between $p$ and $q$ and (3) the total variation of the state is a sum of local variations $\dd=\sum_p\dd_p$. This decomposition is non-canonical in the sense that it requires chosing an assignment of variations of parameters to sites.
From this we define the (non-canonical) notion that the Berry
curvature at site $p$ as $F^{(2)}_p=\dd_p\mc{A}$. The flow of Berry curvature from site $p$ to site $q$ is then some quantity $F^{(3)}_{pq}$, that satisfies the continuity equation, known also in this context as the descent equation\cite{KS2020_higherberry,KS2020_higherthouless}
\begin{equation}
	\sum_{p}F^{(3)}_{pq}=\dd F_{q}^{(2)}.
	\label{eq:descent}
\end{equation}
Naturally this flow must be antisymmetric in $p$ and $q$ as well as a differential $3$-form.
Using the derivative operators $\dd_p$ an immediate solution is
\begin{equation}
	F^{(3)}_{pq}=\dd_p\dd_q\mc{A}.
\end{equation}
This is the higher Berry flow, and is not uniquely specified by equation \eqref{eq:descent}, but this non-uniqueness does not affect the curvature.
Then the higher Berry curvature will be the net flow from $A$ to $B$\cite{KS2020_higherberry}
\begin{equation}
	\Omega^{(3)}=F^{(3)}_{AB}=\sum_{p< a<q}F_{pq}^{(3)}.
\end{equation}
Concretely we can construct such a set of local derivative operators $\dd_p$ by using left canonical MPS which are related to Schmidt decomposition\cite{2003Vidal}.The reader is directed towards the reviews \onlinecite{ORUS2014117} and \onlinecite{2020RMP} for further details on MPS. See the appendix for details on our notational conventions.
\subsection{Higher Berry Curvature from MPS}
Let us define a state by specifying a left canonical MPS representation with diagonal right environments $|r_p)$, and identity left environments $(l_p|$.
Taking the derivative operator $\dd_p$ to act only on the tensor associated with site $p$, this will be the local derivative operator we sought. It is straightforward to evaluate the Berry curvature at site $p$
\begin{equation}
	\nonumber{
		F_{p}^{(2)}=\sum_{r<p}\Im(l_r|\bb{E}^{\dd A_r}_{A_r}\bb{E}(r\to p)\dd \bb{E}^{A_p}_{A_p}|r_p)+\Im(l_p|\bb{E}^{\dd A_p}_{\dd A_p}|r_p)}
\end{equation}
Likewise the flow of Berry curvature from point $p$ to point $q>p$ is
\begin{align}
	F_{pq}^{(3)} & =-\sum_{r<p}\Im(l_r|\bb{E}^{\dd A_r}_{A_r}\bb{E}(r\to p)\dd\bb{E}^{A_p}_{A_p}\bb{E}(p\to q)\dd\bb{E}_{A_q}^{A_q}|r_q)\nonumber \\
	             & \phantom{==\,}-\Im(l_p|\bb{E}^{\dd A_p}_{\dd A_p}\bb{E}(p\to q)\dd\bb{E}^{A_q}_{A_q}|r_q).
	\label{eq:F_pqMPS}
\end{align}
Remarkably, by the relation between the MPS and the Schmidt states\cite{2003Vidal} the higher Berry curvature of the partition between $A$ and $B$, $\Omega^{(3)}=F_{AB}^{(3)}=\sum_{p<a<q}F_{pq}^{(3)}$, is given by equation \eqref{eq:schmidt_hbc_intro}.
From this, it easy to see that $\Omega^{(3)}$ is independent of the choice of left canonical MPS with diagonal right environment, since the remaining gauge freedom corresponds to an overall phase, and a rotation among degenerate Schmidt states. Within such a degenerate sector, the higher Berry curvature is just the coefficient $\dd c^2_\alpha$ times the trace of the conventional non-Abelian Berry curvature of the sector. Because of this gauge freedom we can work in left tangent space gauge\footnote{The tangent space of the manifold of MPS $\bb{A}$ can be decomposed into horizontal and vertical components $T_{A}\bb{A}=H_A\oplus V_A$, where the vertical component corresponds to the additive gauge freedom, arising from the choice basis on the virtual space in the MPS. The choice of horizontal component $H_A$ is not canonical, but one choice is the left tangent space condition where the tangent vector $B$ obeys $(l_p|\bb{E}^{B}_A=0$. Since we verify that $\Omega^{(3)}$ is gauge invariant, i.e. horizontal, we can may impose the left tangent space condition directly on this expression.}\cite{2019vanderstaeten}, i.e. $(l_p|\bb{E}^{\dd A_p}_{A_p}=0$, and a short computation shows that $\Omega^{(3)}$ can be expressed in terms of MPS tensors as
\begin{align}
	\Omega^{(3)}
	 & =-\Im\sum_{p<a}(l_p|\bb{E}_{\dd A_p}^{\dd A_p}\bb{E}(p\to a)|\dd r_a)\label{eq:MPS_HBC} \\
	 & =-\Im\sum_{p<a}
	\begin{tikzpicture}[scale=1,
			baseline ={([yshift=-2]current bounding box.center)},
			inner sep =1.5mm]
		\node[tensor] (top) at (0,0.5) {$\dd A_p$};
		\node[tensor] (bot) at (0,-0.5) {$\dd\bar{A_p}$};
		\draw[-] (top)--(bot);
		\node[minimum size=10] (proptopleft) at (1,0.5) {$\,$};
		\node[minimum size=10] (proptopright) at (2.5,0.5) {$\,$};
		\node[minimum size =10] (propbotleft) at (1,-0.5) {$\,$};
		\node[minimum size =10] (propbotright) at (2.5,-0.5) {$\,$};
		\node[tensorcirc,minimum size=0] (r) at (3.5,0.0) {$\dd r_a$};
		\draw[-] (proptopleft.135) rectangle (propbotright.315) node[pos=.5]
			{$\mathbb{E}(p\to a)$};
		\draw [-](top) -- (proptopleft);
		\draw [-](bot) --(propbotleft);
		\draw[-] (bot.west) .. controls +(-0.5, 0) and +(-0.5, 0) .. (top.west);
		\draw[-] (propbotright.east) .. controls +(0.25, 0) and +(0.0, -0.25) .. (r.south);
		\draw[-] (proptopright.east) .. controls +(0.25, 0) and +(0.0, +0.25) .. (r.north);
	\end{tikzpicture}.\label{eq:MPS_HBC_diagram}
\end{align}
If we do not impose the left tangent space gauge condition, then there is an additional term of the form:
\begin{equation}
	-\Im\sum_{p<q<a}(l_p|\bb{E}^{\dd A_p}_{A_p}\bb{E}(p\to q)\dd \bb{E}^{A_q}_{ A_q}\bb{E}(q\to a)|\dd r_a)\label{eq:anomalous}
\end{equation}
corresponding to the derivatives acting on two different sites on the left side of the cut.
So far, we have worked with finite systems, where the integral $\int_X \Omega^{(3)}$ over a closed $3$-manifold $X$ must vanish because we are not distinguishing between the bulk and the edge. The higher Berry curvature is then the total derivative of the amount of regular curvature on the left part of the system $\Omega^{(3)}=\dd\left(\sum_{\alpha}c_{\alpha}^2\braket{\dd\alpha|\dd\alpha}_{[A]}\right)=\dd F^{(2)}_A$, so its integral is $\int_X\Omega^{(3)}=\int_{\partial X}F_{A}^{(2)}$.
This triviality argument may be circumvented by keeping track of the edges (so $\partial X\neq0$). Alternatively note that the MPS expression works equally well in infinitely large systems, and since $F_{A}$ is extensive it is ill defined, and the edges are then sharply distinguished from the bulk.
We are often interested in
systems with ground states having translational invariance, where we can express the state with a single matrix $A^s$,
which is known as the {uniform MPS} (uMPS)\cite{2020RMP}.
Then equation \eqref{eq:MPS_HBC} can be efficiently evaluated (in the left tangent space gauge).
\begin{align}
	\Omega^{(3)} & =-\Im(l|\bb{E}^{\dd A}_{\dd A}(1-\bb{E}^{A}_{A})^{-1}|\dd r), \\
	\label{eq:uMPS_HBC}
	             & =  -\Im
	\begin{tikzpicture}[scale=1,
			baseline ={([yshift=-2]current bounding box.center)},
			inner sep =1.5mm]
		\node[tensor] (top) at (0,0.5) {$\dd A$};
		\node[tensor] (bot) at (0,-0.5) {$\dd\bar{A}$};
		\draw[-] (top)--(bot);
		\node[minimum size=10] (proptopleft) at (1,0.5) {$\,$};
		\node[minimum size=10] (proptopright) at (2.5,0.5) {$\,$};
		\node[minimum size =10] (propbotleft) at (1,-0.5) {$\,$};
		\node[minimum size =10] (propbotright) at (2.5,-0.5) {$$};
		\node[tensorcirc,minimum size=0] (r) at (3.5,0.0) {$\dd r$};
		\draw[-] (proptopleft.135) rectangle (propbotright.315) node[pos=.5]
			{$\left(1-\mathbb{E}^A_A\right)^{-1}$};
		\draw [-](top) -- (proptopleft);
		\draw [-](bot) --(propbotleft);
		\draw[-] (bot.west) .. controls +(-0.5, 0) and +(-0.5, 0) .. (top.west);
		\draw[-] (propbotright.east) .. controls +(0.25, 0) and +(0.0, -0.25) .. (r.south);
		\draw[-] (proptopright.east) .. controls +(0.25, 0) and +(0.0, +0.25) .. (r.north);
	\end{tikzpicture}
\end{align}
again if we do not impose the left tangent space gauge, there is the additional contribution:
\begin{equation}
	-\Im(l|\bb{E}^{\dd A}_A(1-\bb{E}^{A}_A)^{-1}\dd \bb E_A^A(1-\bb{E}^{A}_A)^{-1}|\dd r)\label{eq:uMPS_additional}.\end{equation}
We can manifestly see that for (essentially \cite{2023Ryu}) injective MPS,
the higher Berry curvature $\Omega^{(3)}$ is convergent due to the normalisation of the state, which is implicit in summing the geometric series.
\subsection{Calculation of Higher Berry curvature in concrete models}
\label{Sec:Toy}

To illustrate the validity of the above approach, we apply it to calculate the higher Berry curvature of the exactly solvable model introduced in Ref.~\onlinecite{qpump2022}. The gerbe structure of this model was studied in\cite{2023Qi}.

Consider a one dimensional lattice $\mathbb Z$, with local Hilbert space $\mc{H}_p=\mathbb{C}^2$ having associated Pauli matrices $\bm{\sigma}_p=(\sigma^1_p,\sigma_p^2,\sigma_p^3)$. We denote the spin $1/2$ coherent state along the direction of any vector $\vb n\in \R^3$ by $\ket{\vb n}$.
We are interested in the family of wave functions $\ket{\psi}$ defined over the parameter space that is the unit sphere $X=S^3\subseteq \R^4$, whose points we label $w=(w_1,w_2,w_3,w_4)=(\vb{w},w_4)$ which may be parameterised by hyperspherical angles $\alpha,\theta\in[0,\pi]$, and $\phi\in[0,2\pi)$\footnote{i.e. $w_4=\cos\alpha$, $\vb w=\sin\alpha \hat{\vb w}$, and $\hat{\vb w}=(\sin\theta\cos\phi,\sin\theta\sin\phi,\cos\theta)$.}. The wave functions are the ground states of the family of Hamiltonians
\begin{equation}
	H_\mr{1d}(w)=\sum_p(-1)^p\vb w\cdot \bm{\sigma}_p+
	g_p(w_4)\bm{\sigma}_p\cdot \bm{\sigma}_{p+1},
	\label{eq:model_ham}
\end{equation}
where the onsite term takes the form of a Zeeman coupling with alternating sign
and the interaction is the antiferromagnetic Heisenberg term whose coefficient depends on $w_4$, as $g_p(w_4)=w_4\delta_{p\in 2\Z+1}\theta(w_4)-w_4\delta_{p\in2\Z}\theta(-w_4)$\footnote{Note that while this Hamiltonian is continuous it is not smooth at $w_4=0$, an equally good model can be obtained by replacing $g_p(w_4)=g_1(w_4)\delta_{p\in 2\Z+1}+g_2(w_4)\delta_{p\in 2\Z}$ for arbitrary $g_{1,2}\geq 0$ such that $g_1(-1)=g_2(1)=0$ and $g_1(w_4)g_2(w_4)=0$ so long as the gap of the Hamiltonian doesn't close.}.
The Hamiltonian dimerizes and so is exactly solvable. It can be visualized for different values of $w_4 \in [-1,1]$ as:
\begin{equation}
	\label{eq:H_config}
	\small
	\begin{tikzpicture}

		\node at (-60pt,0pt){$0<w_4<1$:};

		\draw [dashed](-10+40pt,30pt)--(-10+40pt,-76pt);

		\notsotiny
		\draw (-20pt,0pt) circle (4.5pt);
		\node at (-20pt,0pt){$+$};
		\draw [thick](4.5-35pt,0pt)--(15.5-40pt,0pt);

		\draw (0pt,0pt) circle (4.5pt);
		\node at (0pt,0pt){$-$};
		\draw [thick](4.5pt,0pt)--(15.5pt,0pt);
		\draw (20pt,0pt) circle (4.5pt);
		\node at (20pt,0pt){$+$};

		\draw (40pt,0pt) circle (4.5pt);
		\node at (40pt,0pt){$-$};
		\draw [thick](44.5pt,0pt)--(55.5pt,0pt);
		\draw (60pt,0pt) circle (4.5pt);
		\node at (60pt,0pt){$+$};

		\draw (80pt,0pt) circle (4.5pt);
		\node at (80pt,0pt){$-$};
		\draw [thick](84.5pt,0pt)--(95.5pt,0pt);
		\draw (100pt,0pt) circle (4.5pt);
		\node at (100pt,0pt){$+$};

		\draw (120pt,0pt) circle (4.5pt);
		\node at (120pt,0pt){$-$};
		\draw [thick](124.5pt,0pt)--(135.5pt,0pt);
		\draw (140pt,0pt) circle (4.5pt);
		\node at (140pt,0pt){$+$};

		\begin{scope}[yshift=-22pt]
			\small
			\node at (-51pt,0pt){$w_4=0$:};

			\notsotiny
			\draw (-20pt,0pt) circle (4.5pt);
			\node at (-20pt,0pt){$+$};

			\draw (0pt,0pt) circle (4.5pt);
			\node at (0pt,0pt){$-$};

			\draw (20pt,0pt) circle (4.5pt);
			\node at (20pt,0pt){$+$};

			\draw (40pt,0pt) circle (4.5pt);
			\node at (40pt,0pt){$-$};

			\draw (60pt,0pt) circle (4.5pt);
			\node at (60pt,0pt){$+$};

			\draw (80pt,0pt) circle (4.5pt);
			\node at (80pt,0pt){$-$};

			\draw (100pt,0pt) circle (4.5pt);
			\node at (100pt,0pt){$+$};

			\draw (120pt,0pt) circle (4.5pt);
			\node at (120pt,0pt){$-$};

			\draw (140pt,0pt) circle (4.5pt);
			\node at (140pt,0pt){$+$};

		\end{scope}

		\begin{scope}[yshift=-44pt]

			\small

			\node at (-63pt,0pt){$-1<w_4<0$:};

			\notsotiny

			\draw (-20pt,0pt) circle (4.5pt);
			\node at (-20pt,0pt){$+$};

			\draw (0pt,0pt) circle (4.5pt);
			\node at (0pt,0pt){$-$};
			\draw [thick](4.5-20pt,0pt)--(15.5-20pt,0pt);
			\draw (20pt,0pt) circle (4.5pt);
			\node at (20pt,0pt){$+$};

			\draw (40pt,0pt) circle (4.5pt);
			\node at (40pt,0pt){$-$};
			\draw [thick](24.5pt,0pt)--(35.5pt,0pt);
			\draw (60pt,0pt) circle (4.5pt);
			\node at (60pt,0pt){$+$};

			\draw (80pt,0pt) circle (4.5pt);
			\node at (80pt,0pt){$-$};
			\draw [thick](64.5pt,0pt)--(75.5pt,0pt);
			\draw (100pt,0pt) circle (4.5pt);
			\node at (100pt,0pt){$+$};

			\draw (120pt,0pt) circle (4.5pt);
			\node at (120pt,0pt){$-$};
			\draw [thick](104.5pt,0pt)--(115.5pt,0pt);
			\draw (140pt,0pt) circle (4.5pt);
			\node at (140pt,0pt){$+$};
			\draw [thick](144.5pt,0pt)--(150.5pt,0pt);
		\end{scope}

		\begin{scope}[yshift=-66pt]

			\draw [thick](-20pt,0pt)--(0pt,0pt);
			\draw [thick](20pt,0pt)--(40pt,0pt);
			\draw [thick](60pt,0pt)--(80pt,0pt);
			\draw [thick](100pt,0pt)--(120pt,0pt);
			\draw [thick](140pt,0pt)--(150pt,0pt);

			\node at (140pt,0pt){$\bullet$};
			\node at (120pt,0pt){$\bullet$};
			\node at (100pt,0pt){$\bullet$};
			\node at (80pt,0pt){$\bullet$};
			\node at (60pt,0pt){$\bullet$};
			\node at (40pt,0pt){$\bullet$};
			\node at (20pt,0pt){$\bullet$};
			\node at (0pt,0pt){$\bullet$};
			\node at (-20pt,0pt){$\bullet$};

			\small
			\node at (-55pt,0pt){$w_4=-1$:};
		\end{scope}

		\begin{scope}[yshift=22pt]

			\draw [thick](-30pt,0pt)--(-20pt,0pt);
			\draw [thick](0pt,0pt)--(20pt,0pt);
			\draw [thick](40pt,0pt)--(60pt,0pt);
			\draw [thick](80pt,0pt)--(100pt,0pt);
			\draw [thick](120pt,0pt)--(140pt,0pt);

			\node at (140pt,0pt){$\bullet$};
			\node at (120pt,0pt){$\bullet$};
			\node at (100pt,0pt){$\bullet$};
			\node at (80pt,0pt){$\bullet$};
			\node at (60pt,0pt){$\bullet$};
			\node at (40pt,0pt){$\bullet$};
			\node at (20pt,0pt){$\bullet$};
			\node at (0pt,0pt){$\bullet$};
			\node at (-20pt,0pt){$\bullet$};

			\small
			\node at (-52pt,0pt){$w_4=1$:};



		\end{scope}


	\end{tikzpicture}
\end{equation}
We use $\pm$ to represent the sign of the Zeeman coupling at this site, while $\bullet$ represents the case of vanishing Zeeman coupling. Interaction terms are represented by solid lines joining pairs of lattice sites.
For $\pi/2\leq \alpha\leq \pi$ the left canonical MPS of this model is
\begin{equation}
	\label{LeftCanonical_main}
	A^{s_p}=\begin{cases}
		(\braket{s_p|\vb w}, -\braket{s_p|-\vb w})       & \qquad p\in 2\Z   \\
		(c_+\braket{s_p|-\vb w},c_-\braket{s_p|\vb w})^T & \qquad p\in 2\Z+1
	\end{cases}
\end{equation}
where $c_\pm =\sqrt{(1\pm \sin \alpha)/2}$ are the Schmidt entanglement coefficients\footnote{This MPS has the particularly simple interpretation that it is the interpolation between the state that minimises the Zeeman interaction $\ket{\vb w}\otimes \ket{-\vb w}$, and the spin singlet $\ket{\vb w}\otimes \ket{-\vb w}-\ket{-\vb w}\otimes \ket{\vb w}$ that minimises the antiferromagnetic interaction.}.
With this MPS representation,
we can calculate the 3-form higher Berry curvature $\Omega^{(3)}$ across the cut at $a=1/2$ which is the dashed line pictured in \eqref{eq:H_config}.
Applying equation \eqref{eq:MPS_HBC} the higher Berry curvature is
\begin{align}
	\Omega^{(3)} & =\theta(\alpha-\pi/2)\Im\left[\dd c_+^2\braket{\dd\vb w|\dd \vb w}+\dd c_-^2\braket{-\dd\vb w|-\dd \vb w}\right]\nonumber \\
	             & =-\theta(\alpha-\pi/2)\frac12\cos\alpha\sin\theta~\dd\alpha\dd\theta\dd\phi.\label{eq:3-form_toy}
\end{align}
The integral of the higher Berry curvature is quantized as expected $\frac 1 {2\pi}\int_{X}\Omega^{(3)}=1$.

\subsubsection*{Numerical computation of higher Berry curvature for $d=1$ MPS}
\label{sec:numerics}
\begin{figure*}
	\centering
	\includegraphics[width=0.9\linewidth]{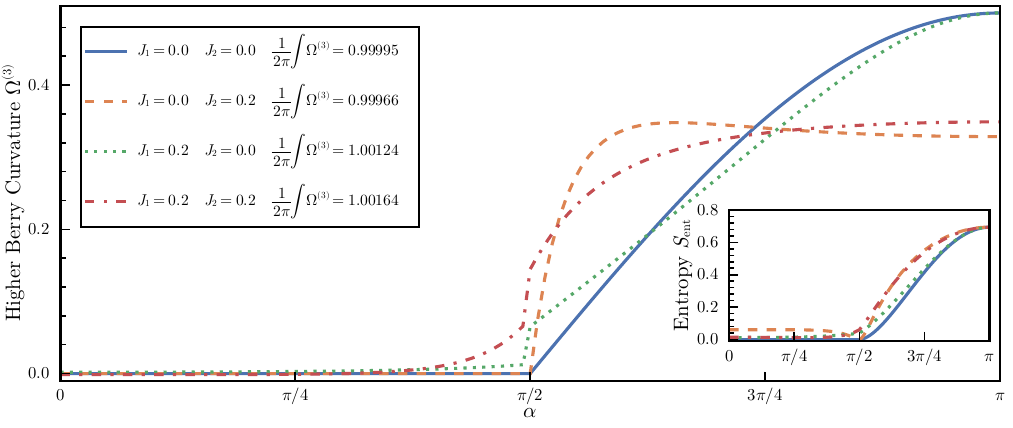}
	\caption{Higher Berry curvature $\Omega^{(3)}$ computed using iDMRG\cite{white1992dmrg,white1993dmrg,mcculloch2008infinite} for a collection of models $H_{J_1J_2}$ as a function of the parameter $\alpha\in[0,\pi]$, with $\theta=\pi/2,\phi=0$. We plot $\Omega_{\alpha\theta\phi}$ which is given as $\Omega^{(3)}=\Omega_{\alpha\theta\phi}\dd\alpha\,\mathrm{dcos}\,\theta\,\dd\phi$. In all cases we see excellent quantization of the integrated Berry curvature to within $0.2\%$. As expected from the analytic calculation, for $J_1=J_2=0$, $\Omega_{\alpha\theta\phi}^{(3)}$
		vanishes for $\alpha\in[0,\pi/2]$ while equalling $-\frac12 \cos\alpha$ for $\alpha\in[\pi/2,\pi]$.
		(Inset:) The entanglement entropy for the half system with the entanglement cut at $a=1/2$.
	}
	\label{fig:numerical_hbc}
\end{figure*}
Consider deforming the model from the previous section as done in Ref.\onlinecite{2023Shiozaki} with additional nearest neighbour and next-nearest neighbour interactions, so that it no longer dimerizes
\begin{equation}
	H_{J_1J_2}(w)=H_\mr{1d}(w)+\sum_p \left[J_1\bm\sigma_p\cdot \bm{\sigma}_{p+1}+J_2\bm{\sigma}_p\cdot\bm{\sigma}_{p+2}\right].\label{eq:non-solvable}
\end{equation}
Using equation \eqref{eq:uMPS_HBC} it is straightforward to calculate the  higher Berry curvature of this deformed model, using for example iDMRG\cite{white1992dmrg,white1993dmrg,mcculloch2008infinite} to find the ground state
and a finite difference approximation to the relevant derivatives.
The resulting higher Berry curvature is plotted in Fig.\ref{fig:numerical_hbc} for the cut $a=1/2$
for four sets of parameters $J_1=0,0.2$ and $J_2=0,0.2$. In particular, fix a discretization of $X$,
and for each point $x$ in this discretization, find via IDMRG the ground state uMPS $A$ of \eqref{eq:non-solvable},
and the ground state at $x+\delta$ for  3 linearly independent small pertubations $\delta$. Then fix a smooth local
gauge about $x$\footnote{To fix the relative gauge of $A$ and the perturbed
$A_\mathrm{perturb}$, we compute the left maximal eigenvector $(v|$ of the mixed transfer
matrix $T=\bb{E}^{A_\mr{perturb}}_A$. To ensure that this a gauge transformation of left canonical MPS, we subtract of all
contributions to $(v|$ which are not block diagonal, with the blocks being the Schmidt degenerate sectors.
To ensure that the gauge transformation is unitary, we perform a QR decomposition in each block, and let the
relative gauge transformation $g$ correspond to just the blockwise unitaries that form the Q-factor.
The phase of the maximal eigenvalue is $z\in \mr{U}(1)$, and we let the gauge fixed perturbed tensor
be $A_\mathrm{gf,perturbed}^s=z g A^s_\mathrm{perturbed}g^\dagger $, and then
$\dd A\simeq \frac{1}{\delta}(A_\mathrm{gf,perturbed}-A)$.}, and take finite differences of the uMPS tensor $A$ and its perturbation. For IDMRG we pick a
Schmidt coefficient tolerance of $10^{-7}$, and the magnitude of the step
$\delta$ in $\theta,\alpha,\phi$ is $10^{-5}$. There is a parameter space symmetry corresponding to rotations
of $\vb{w}$, so we only need to sample $\alpha$, and for $201$ sample points we find quantization
of the higher Berry invariant to within $0.2\%$.
While important for the topological properties of gerbes which follow, in practical calculations, the distinction between essentially injective and injective MPS is immaterial. When the Schmidt rank is constant we are free to use injective MPS to calculate the curvature and the Schmidt rank only changes in measure 0 region, which cannot be detected in our numerics.
We implement our computation in Julia\cite{bezansonJuliaFreshApproach2017} using MPSKit\cite{maartenvdMPSKitJl2023}.

\section{Quantization of the
  higher Berry curvature of uMPS}

\label{sec:gerbe}
We show explicitly that our 3-form higher Berry curvature for uMPS correspond to the 3-form curvature of a gerbe\cite{1999Hitchin}, and so integrates to values in $2\pi\Z$, by providing explicitly the $1$- and $2$-connections in terms of uMPS, and showing they satisfy the de Rahm-\v{C}ech descent equations.

	{\it Review of Gerbes:} Gerbes are topological structures over $X$ as a generalization of line bundles, which can be specified in terms of transition functions over some open cover in an analogous way\cite{1999Hitchin}
The essentially/semi injective uMPS over a parameter manifold $X$ form a gerbe\cite{2023Shiozaki,2023Ryu}, and we chose to follow the construction in\cite{2023Shiozaki}, which we briefly review. Choosing a good open cover $\{U_{\alpha}\}$ of $X$ we specify essentially injective smooth tensors $A^{s}_{\alpha}$ on each open set $U_{\alpha}$, which are taken to be left-canonical with a normalised diagonal right environment $|r_{\alpha})$ which when viewed as an operator will be denoted $r_{\alpha}$, with injective part $\tilde{r}_\alpha$.
On double overlaps $U_{\alpha\beta}=U_{\alpha}\cap U_{\beta}$, the mixed transfer matrix $\bb{E}^{A_{\beta}}_{A_{\alpha}}$ has maximal left eigenvector $(l_{\alpha\beta}|$,
which when viewed as an operator on the virtual space will denoted $l_{\alpha\beta}$. Up to an overall phase, the injective part of the essentially injective MPS tensors $A_\alpha$ $A_\beta$ will be related by the conjugation of a unitary transformation $g_{\alpha\beta}$ equal to the injective part of $l_{\alpha\beta}$. Since $\tilde{r}_{\alpha}$ is the squared Schmidt coefficients, which are canonical up to permutation, we have $\dd g_{\alpha\beta} \tilde{r}_\beta =\tilde{r}_{\alpha}\dd g_{\alpha\beta}$.
The Dixmier-Douady class in $H^{3}(X,\Z)\simeq H^{2}(X,\underline{\mr{U}(1)})$ has representative in the \v{C}ech cohomology given by $c_{\alpha\beta\gamma}=\tr r_{\alpha}l_{\alpha\beta}l_{\beta\gamma}l_{\gamma\alpha}$ on triple overlaps $U_{\alpha\beta\gamma}=U_\alpha\cap U_{\beta}\cap U_\gamma$.

The curvature $\Omega^{(3)}$ of a gerbe can be related to the \v{C}ech cocycle using the descent equations in the de Rahm-\v{C}ech complex\cite{1999Hitchin,Murray_2007}. Thus we must specify the $2$-form connection $F_\alpha$ on $U_{\alpha}$, a $1$-form connection $B_{\alpha\beta}=-B_{\beta\alpha}$ on $U_{\alpha\beta}$, along with $\Omega^{(3)}$ and $c_{\alpha\beta\gamma}$, satisfying compatibility conditions:
\begin{align}
	B_{\alpha\beta}+B_{\beta\gamma}+B_{\gamma\alpha} & =-\i c_{\alpha\beta\gamma}^{-1}\dd c_{\alpha\beta\gamma},\label{eq:cechAC} \\
	F_{\alpha}-F_\beta                               & =\dd B_{\alpha\beta},\label{eq:cechFA}                                     \\
	\Omega^{(3)}                                     & = \dd F_\alpha.\label{eq:cechOmegaF}
\end{align}
If $\Omega^{(3)}$ satisfies these conditions, and $c_{\alpha\beta\gamma}$ is a \v{C}ech cocycle, the integral $\int_{X}\Omega^{(3)}\in 2\pi\Z$.\\
\textit{uMPS connective structure:}
The $3$ form curvature $\Omega^{(3)}$ is as in \eqref{eq:uMPS_HBC} and \eqref{eq:uMPS_additional}, the $1$-connection is
\begin{equation}
	B_{\alpha\beta}=-\i \tr(r_{\beta}l_{\beta\alpha}\dd l_{\alpha\beta})\label{eq:A12}
\end{equation}
and the $2$-connection is
\begin{equation}
	F_{\alpha}=-\i \left(l_\alpha\left|\bb{E}^{\dd A_{\alpha}}_{A_\alpha}\left(1-\bb{E}_{A_{\alpha}}^{A_\alpha}\right)^{-1}\right|\dd r_\alpha\right).
\end{equation}
Remarkably the $2$-connection is equal to the overlap of $|\dd r)$ and the left connection of the bundle of gauge equivalent uMPS tensors\cite{Haegeman_2014}.
Proving that equation \eqref{eq:cechAC} is satisfied is straightforward by noting that $\tr\dd r_\alpha l_{\alpha\beta}l_{\beta\gamma}l_{\gamma\alpha}=\tr \dd \tilde{r}_{\alpha} g_{\alpha\beta}g_{\beta\gamma}g_{\gamma\alpha}=c_{\alpha\beta\gamma}\tr\dd \tilde{r}_{\alpha}=0$.
Since only the injective part contributes we find
$c_{\alpha\beta\gamma}^{-1}\dd c_{\alpha\beta\gamma}=\tr r_{\alpha}\dd l_{\alpha\beta}l_{\beta\alpha}+\text{(cyclic permutations)}$.
To relate the $1$- and $2$-connection, we evaluate
\begin{equation}
	\dd B_{\alpha\beta}=-\i \tr\dd l_{\beta\alpha}\dd l_{\alpha\beta}r_{\beta}+\i\tr l_{\beta\alpha}\dd l_{\alpha\beta}\dd r_{\beta}
\end{equation}
Since we are freely able to project onto the injective part inside the trace, due to the presence of the (derivative of) right environment, and $g_{\alpha\beta}g_{\beta\alpha}=1$, the first term vanishes. From the defining eigenvalue equation
\begin{equation}
	\left(\dd l_{\alpha\beta}\right|=C\left(l_{\alpha\beta}\right|
	+\left(l_{\alpha\beta}\right|\dd\bb{E}^{A_{\beta}}_{A_{\alpha}}\left(1-\bb{E}^{A_{\beta}}_{A_{\alpha}}\right)^{-1}
\end{equation}
where $C$ is some one-form that cannot be determined from the eigenvalue equation, but drops out of the overlap. Now $\dd \bb{E}^{A_\beta}_{A_\alpha}=\bb{E}^{\dd A_\beta}_{A_{\alpha}}+\bb{E}^{A_\beta}_{\dd A_\alpha}$, and since $\tilde{r}_{\alpha,\beta}$ differ only by a permutation of values $\dd B_{\alpha\beta}=F_{\alpha}-F_{\beta}$
as promised. Finally $\dd F_\alpha$ has two terms, one where the derivative acts on $\bb{E}^{\dd A_{\alpha}}_{A_\alpha}$, which gives \eqref{eq:uMPS_HBC}, and one where the derivative acts on $(1-\bb{E}_A^A)^{-1}$, yielding \eqref{eq:uMPS_additional}, thus $\Omega^{(3)}=\dd F_{\alpha}$.

\section{Conclusion}
In this paper we have presented a novel approach to defining a local notion of Berry curvature, its flow, and the Higher Berry curvature $\Omega^{(3)}$ for wave functions parameterised with MPS. Gerbes are the generalizations to $d=1$ of the line bundles of $d=0$ quantum mechanical systems.
We have shown that $\Omega^{(3)}$ is the curvature of the gerbe of essentially injective uMPS, and hence has quantized integrals over closed parameter spaces. Our main insight was to generalise the notion of local variation beyond varying the parent Hamiltonian, in this case to a MPS parameterization. Our approach readily generalizes to tensors in higher dimensions, and we can construct the higher Berry curvature as well as the higher Thouless pumps in the presence of $\mr{U}(1)$ symmetry\cite{2024long}.
In a finite system with edges, our expression for the HBC is still expected to be quantized up to corrections exponentially small in the ratio of the correlation length to the system size, if contributions from the edge are excluded. Alternatively edges can be included, if we allow them to accumulate net Chern number.
In the non-uniform case $\Omega^{(3)}$ depends on the choice of the cut, corresponding to the fact that the flow need not be uniform, but the quantized invariant is independent of this choice. We expect a similar situation arises upon changing the non-canonical choice of local Berry curvature $F_{p}^{(2)}$, which explains why the curvature $\Omega^{(3)}$ of the exactly solvable model differs between this paper and the Hamiltonian calculation of \cite{qpump2022}, but the invariant is the same.
Some future directions of interest follow:
As elaborated in \cite{Kapustin2201}, the equivariant extension of the higher Berry curvature gives rise to the (regular, higher, and nonabelian) Hall conductivity, and it would be interesting to make this connection explicit in tensor networks. It would also be interesting to generalise the uMPS gerbe structure to non-translationally invariant state, and see if the curvature we have defined is the corresponding gerbe curvature. Finally, it would be interesting to study the experimental implications of the higher Berry curvature, and how to detect them.

\bigskip
\textit{Note added}: While completing this manuscript, we became aware of an upcoming related work Ref.\onlinecite{2024Ryu} to appear on arXiv on the same day.

\bigskip

\acknowledgments

We thank Michael Hermele and Cristian Batista for helpful discussions and general comments and Daniel Parker for insights into MPS.
This work was supported in part by the Simons Collaboration on Ultra-Quantum Matter, which is a grant from the Simons Foundation (618615, XW, AV; 651440, XW, MH).

\appendix
\section{MPS notation}
\label{app:mpsnotation}
For an MPS we associate to each link between sites $p$ and $p+1$ a virtual Hilbert space $\mc{H}^\mr{virt}_p$, with orthonormal basis $|\alpha_p))$ where $\alpha_p$ labels Schmidt sectors, and naturally $\mc{H}^\mr{virt}_0=\mc{H}^\mr{virt}_N=\bb{C}$. Then the MPS consists of maps $A_p:\mc{H}^\mr{virt}_p\to \mc{H}_{p-1}^\mr{virt}\otimes \mc{H}_p$, which is expressed as $A|\alpha_p))=\sum_{s_p\alpha_{p-1}}A^{s_p}_{\alpha_{p-1}\alpha_{p}}|\alpha_{p-1}))\otimes\ket{s_p}$. We can write the state as
\begin{align}
	\ket{\psi} & =\sum_{\set{s_p}} A_1^{s_1}\cdots A_N^{s_{N}}\ket{s_1\cdots s_N} \\
	           & =
	\begin{tikzpicture}[
			baseline ={([yshift=9]current bounding box.center)},
			inner sep=1mm]
		\node[tensor] (top) at (0,0.5) {$A_1$};
		\draw[-] (top)-- (top.east)-- +(0.3,0);
		\draw[-] (top.south)+(0,-0.3)--(top);
		\node[tensor] (top) at (1.0,0.5) {$A_2$};
		\draw[-] (top.west)+(-0.3,0)--(top)-- (top.east)-- +(0.3,0);
		\draw[-] (top.south)+(0,-0.3)--(top);
		\node[tensor] (top) at (3.0,0.5) {$A_N$};
		\draw[-] (top.west)+(-0.3,0)--(top);
		\draw[-] (top.south)+(0,-0.3)--(top);
		\node at (0,-0.3) {$s_1$};
		\node at (1,-0.3) {$s_2$};
		\node at (3,-0.3) {$s_N$};
		\node at (2,0.5) {$\cdots$};
		\node at (2,-0.3) {$\cdots$};
	\end{tikzpicture}
\end{align}
where we have used the conventional tensor network contraction diagrams. For a vector space $V$, let $V^\star$ be its dual, and $\bar{V}$ its complex conjugate. From the Hilbert space structure on the physical space we find isomorphism $\overline{\mc{H}_p}\leftrightarrow \mc{H}_p^\star$, so we can define the complex conjugate map $\bar{A}_p:\overline{\mc{H}^\mr{virt}_p}\to\overline{\mc{H}^\mr{virt}_{p-1}}\otimes \mc{H}_p^\star$. There is a natural and useful notion of the transfer matrix between any MPS tensors $A$ and $B$
\begin{equation}
	\bb{E}^A_B=
	\begin{tikzpicture}[
			baseline ={([yshift=0]current bounding box.center)},
			inner sep=1mm]
		\node[minimum size=15] (top) at (0,0.5) {$\,$};
		\node[minimum size =15] (bot) at (0,-0.5) {$\,$};
		\draw[-] (top.135) rectangle (bot.315) node[pos=.5] {$\bb{E}^A_B$};
		\draw[-] (top.west)+(-0.25,0)--(top)-- (top.east)-- +(0.25,0);
		\draw[-] (bot.west)+(-0.25,0)--(bot)--(bot.east)-- +(0.25,0);
	\end{tikzpicture}
	=
	\begin{tikzpicture}[
			baseline ={([yshift=0]current bounding box.center)},
			inner sep=1mm]
		\node[tensor] (top) at (0,0.5) {$A$};
		\node[tensor] (bot) at (0,-0.5) {$\bar{B}$};
		\draw[-] (top)--(bot);
		\draw[-] (top.west)+(-0.25,0)--(top)-- (top.east)-- +(0.25,0);
		\draw[-] (bot.west)+(-0.25,0)--(bot)--(bot.east)-- +(0.25,0);
	\end{tikzpicture}
	\label{eq:TransferMatrix_1d}
\end{equation}
This can be considered a linear operator on the doubled virtual space $\mc{H}_{p}^\mr{virt}\otimes\overline{\mc{H}^\mr{virt}_{p}}\to \mc{H}^\mr{virt}_{p-1}\otimes \overline{\mc{H}^\mr{virt}_{p-1}}$, taking $\bb{E}^{A}_{B}:|\alpha_{p}))\otimes|\overline{\beta_p}))\mapsto \tr_{\mc{H}_p}[A|\alpha_p))\otimes \bar{B}\overline{|\beta_p))}]$. Expectation values of states correspond to pairings of this doubled virtual space and its dual, so for convenience we denote any vector in this space like $|x)$ for some label $x$. For a given MPS denote by $\bb{E}(b\to a)$ the transfer matrix product $\bb{E}(b\to a)=\prod_{b\leq p<a}\bb{E}_{A_q}^{A_q}$. The left environment at site $p$ is by definition $(l_p|=\bb{E}(1\to p)$ and the right environment is $|r_p)=\bb{E}(p+0^+\to N+1)$. Given a state $\ket{\psi}$ we can construct the tensors corresponding to a particularly nice representation. Let the Schmidt decomposition between sites $p$ and $p+1$ of the state be $\ket{\psi}=\sum_{\alpha_p}c_{\alpha_{p}}\ket{\alpha_{p}}_{[1\cdots p]}\ket{\alpha_{p}}_{[p+1\cdots N]}$. Then we define the tensor map $A^{s_p}=\braket{\alpha_{p-1}s_p|\alpha_{p}}_{[1\cdots p]}$ .
For this particular choice of tensors it is clear that $(l_p|=\sum_{\alpha_{p-1}}((\alpha_{p-1}|\otimes\overline{ (({\alpha}_{p-1}|}$ and $|r_p)=\sum_{\alpha_p}c_{\alpha_p}^2|\alpha_p))\otimes\overline{|\alpha_p))}$. When imposing a Hilbert space structure on the virtual spaces as mentioned, we can identify elements of the doubled space as linear maps of the virtual space. Then the left environment is the identity operator, and the right is diagonal and acts by multiplication of the square corresponding Schmidt coefficient. More broadly the tensors need not have been derived from such a Schmidt decomposition, if their environments take this form the MPS is called a left canonical Matrix product state, and any canonical MPS has the aforementioned relationship to the Schmidt decomposition it defines. By keeping only Schmidt sectors with nonzero $c_\alpha$, these MPS are injective, under the assumption that there is a finite correlation length $\xi$, but we will need the slightly broader notion where we allow such zero Schmidt sectors, known as essentially injective\cite{2023Ryu} to have nontrivial higher Berry curvature. As long as we contract with (derivatives of) right environments, this loss of injectivity will not affect observables.
\\

\section{Smooth essentially injective canonical MPS on open covers}
\label{Appendix:Smooth_MPS}

In this section we show how to construct smooth essentially injective uMPS on an open cover of the pump over $X=S^3$ described in the paper. This will require a small thickening of the $\alpha=0,\pi$ poles, as we will describe. Pick arbitrary $\delta,\epsilon$ such that $0<\delta<\epsilon< \pi/2$. Let the parameterization of the three sphere be $\psi:(\alpha,\theta,\phi)\mapsto(w_1,w_2,w_3,w_4)$, then
the open cover will consist of two patches. In particular, consider the closed sets given by the curves $K_\pi=\psi([\epsilon,\pi-\delta]\times\set{\pi}\times[0,2\pi))$, and $K_0=\psi([\epsilon,\pi-\delta]\times\set{0}\times [0,2\pi))$. The open sets will be their complements
$U_\pi=S^3\setminus K_\pi$ and $U_0=S^3\setminus K_0$, and notably their intersection is homotopic to the two-sphere $U_\pi\cap U_0\overset{\mr{homo}}{\simeq} S^2$.
To specify the MPS, we split these opens sets into four components, corresponding to $\alpha\in[0,\epsilon]$, $\alpha\in[\epsilon,\pi-\epsilon]$, $\alpha\in [\pi-\epsilon,\pi-\delta]$ and $\alpha\in[\pi-\delta,\pi]$, and denote these component the north pole $U_{0,\pi}^N$, the regular parts $U_{0,\pi}^\mr{reg}$, the near-south pole $U_{0,\pi}^{S\epsilon}$ and the south pole $U_{0,\pi}^{S}$.
For convenience let $\tilde\alpha = \frac{\pi(\alpha-\epsilon)}{\pi-2\epsilon}$.
On the regular parts of the patches $U^\mr{reg}_{0,\pi}$, we may construct spin coherent states of $\vb w=(w_1,w_2,w_3)$.
In particular $\ket{\vb w}=V_{0,\pi}\ket{\uparrow}$, and $\ket{-\vb w}=V_{0,\pi}\ket{\downarrow}$, and $V_{0,\pi}$ is a smooth unitary away from $\theta=0,\pi$ respectively.
In particular in the $\ket{\uparrow},\ket{\downarrow}$ basis we take, and
\begin{align}
	V_\pi & =\begin{pmatrix}
		         \cos\theta/2              & -\e^{-\i\phi}\sin \theta/2 \\
		         \e^{\i\phi} \sin \theta/2 & \cos\theta/2
	         \end{pmatrix} \\
	V_0   & =\begin{pmatrix}
		         \e^{-\i\phi}\cos\theta/2 & -\sin\theta/2           \\
		         \sin\theta/2             & \e^{\i\phi}\cos\theta/2
	         \end{pmatrix}.
\end{align}
The MPS factorises as a map first to just the north pole and then a spin rotation (though not one smoothly connected to the identity on $U_{0}$).
On $U_{0,\pi}^\mr{reg}$ let the MPS be
\begin{widetext}
	\begin{equation}
		A={V}_{0,\pi}^{\otimes 2}\begin{pmatrix}
			c_+(\tilde\alpha)\ket{\downarrow\uparrow} -d(\tilde \alpha)\ket{\uparrow\downarrow} & (V^{\otimes 2})^{-\sqrt{2}d(\tilde\alpha)}(-c_+(\tilde\alpha)\ket{\downarrow\downarrow}+d(\tilde\alpha)\ket{\uparrow\uparrow}) \\
			c_-(\tilde\alpha)\ket{\uparrow\uparrow}                                             & -c_-(\tilde\alpha)\ket{\uparrow\downarrow}
		\end{pmatrix}
	\end{equation}
\end{widetext}
where $c_\pm$ and $d$ are any sufficiently smooth positive functions such that $c_-$ has support on $[\pi/2,\pi]$, $d$ has support on $[0,\pi/2]$, $d^2+c_-^2+c_+^2=1$ everywhere and $c_+(0)=c_+(\pi)=1/\sqrt{2}$. Specifying $c_+(\tilde\alpha)$ we find $d(\tilde\alpha)=\theta(\pi/2-\tilde\alpha)\sqrt{1-c^2_+(\tilde\alpha)}$ and $c_-=\theta(\tilde\alpha-\pi/2)\sqrt{1-c_+^2(\tilde\alpha)}$.

The original Hamiltonian does not result in a smooth MPS at $\alpha=\pi/2$, but would correspond to the choice $c_+(\tilde\alpha)=\sqrt{\frac{1+\sin\tilde\alpha}{2}}$. We can make the MPS have a continuous derivative everywhere by instead letting for example $c_+(\tilde\alpha)=\sqrt{1-\frac{1}{8}(1+\cos2\tilde\alpha)^2}$ or picking another choice so that the derivatives vanish at $\tilde\alpha=0,\pi/2,\pi$. Now we discuss how to handle the thickened poles. First, consider the North pole near $\alpha=0$, and let
\begin{equation}
	A=\frac{1}{\sqrt 2}\begin{pmatrix}
		\ket{\downarrow\uparrow}-\ket{\uparrow\downarrow} & \ket{\uparrow\uparrow}-\ket{\downarrow\downarrow} \\
		0                                                 & 0
	\end{pmatrix}
\end{equation}
on the entirety of $U_{0,\pi}^N$. If the coefficients $c_\pm,d$ are chosen to have vanishing derivatives to sufficient order at $\tilde\alpha=0$, this is not just continuous but also smooth. For $\alpha=\pi-\epsilon$, $A$ on $U_{0,\pi}$ is gauge equivalent to
\begin{equation}
	A_S=\frac{1}{\sqrt{2}}\begin{pmatrix}
		\ket{\downarrow\uparrow} & -\ket{\downarrow\downarrow} \\
		\ket{\uparrow\uparrow}   & -\ket{\uparrow\downarrow}
	\end{pmatrix}
\end{equation}
by a unitary transformation $g$ such that $A^s(\alpha=\pi-\epsilon)=g^\dagger A^s_S g$, which is exactly equal to $V_{0,\pi}$ acting on the virtual indices. For $\alpha\in[\pi-\epsilon,\pi-\delta]$. Let $x(\alpha)$ be a smooth function such that $x(\pi-\delta)=0$, $x(\alpha-\epsilon)=1$ and which has vanishing derivativies to all orders at these points. Then define
$A^s(\alpha)=(g^\dagger)^xA^s_Sg^x$ on $U_{0,\pi}^{S\epsilon}$. For $\alpha\in[\pi-\delta,\pi]$ and any $\theta,\phi$, we let $A=A_S$, which defines it on $U^S_{0,\pi}$.

The Schmidt coefficients are constant on the enlarged poles, so they contribute no curvature, while on the regular part of $S^{3}$, we find the curvature is $0$ for $0<\alpha<\pi/2$, and $\Omega^{(3)}=-\sin\theta\, \dd c_+^2(\tilde\alpha)\,\dd\theta\,\dd \phi$ for $\pi/2<\alpha<\pi-\epsilon$, so the integrated higher Berry curvature is $\int_X \Omega^{(3)}=4\pi[ c_+^2(\pi/2)-c_+^2(\pi)]=2\pi$.
Noting that the intersection $U_{0}\cap U_\pi$ contracts to the $2$-sphere $S^2$ defined by $\theta=\pi/2$,
let us construct a cellulation of $S^3$, with two $3$-cells corresponding to each of $U_{0,\pi}$, while there is a $2$-cell $Y^2$ corresponding to the sphere $\theta=\pi/2$. While it is not possible to find a globally smooth left eigenvector of the transfer matrix, for the $2$-connections we need only know it when the right environment is changing which occurs on $U_0^\mr{reg}$. Here the left environment is
\begin{equation}
	\Lambda^L_{12}=\begin{pmatrix}
		\e^{\i\phi} & 0 \\0&\e^{-\i\phi}
	\end{pmatrix}
\end{equation}
hence from equation \eqref{eq:A12} we find the difference in $2$-connections on $Y^2_\mr{reg}=Y^2\cap U_0^\mr{reg}$, is
\begin{equation}
	F_{Y}=F_1-F_2=-\frac{1}{2}\dd c_+^2(\tilde\alpha)\dd \phi
\end{equation}
which is the curvature on $Y^2$. Thus $Y^2$ has Chern number
\begin{equation}
	\frac{1}{2\pi}\int_{Y^2} F_Y=1
\end{equation}
consistent with the higher Berry curvature.

\section{Two-site calculation}
In this appendix we briefly illustrate how to compute the Berry curvature flow for a two site model, with sites $A$ and $B$, and Hilbert spaces $\mc{H}_A$ and $\mc{H}_B$, so that the wave function lives in the Hilbert space $\mc{H}=\mc{H}_A\otimes\mc{H}_B$. Let $\ket{\psi}$ be this wave function, and with Schmidt decomposition $\ket{\psi}=\sum_\alpha c_\alpha\ket{\alpha}_{[A]}\ket{\alpha}_{[B]}$. We define the local parameter space of site $A$ to be $X_A=\{\ket{\alpha}_{[A]}\}$ while the local parameter space of site $B$ is $X_B=\{\ket{\alpha}_{[B]},c_\alpha\}$.  The connection $1$-form is
\begin{equation}
	\mc{A}=\i\sum_\alpha\left[c_\alpha^2\braket{\alpha|\dd\alpha}_{[A]}+c_\alpha^2\braket{\alpha|\dd\alpha}_{[B]}\right]
\end{equation}
The curvature at site $A$ is
\begin{equation}
	F_A^{(2)}=\i \sum_\alpha c_\alpha^2\braket{\dd\alpha|\dd\alpha}_{[A]}
\end{equation}
while at site $B$ it is
\begin{equation}
	F_B^{(2)}=\i\sum_\alpha\left[\dd c_\alpha^2\braket{\alpha|\dd\alpha}_{[A]}+\dd c_\alpha^2\braket{\alpha|\dd\alpha}_{[B]}+c^2_\alpha\braket{\dd\alpha|\dd\alpha}_{[B]}\right]
\end{equation}
The asymmetry clearly reflects the non-symmetric choice of local parameter spaces. The flow of Berry curvature from $A$ to $B$ is
\begin{equation}
	F_{AB}^{(3)}=\dd_AF_B^{(2)}=-\dd_B F_A^{(2)}=\i \sum_\alpha
	\dd c_\alpha^2\braket{\dd\alpha|\dd\alpha}_{[A]}
\end{equation}
It it apparent that this choice of local parameter space corresponds to the left canonical MPS. For example, let us consider states over $S^3$, similarly to the pump. Then we might consider interpolating $\ket{\vb w}\otimes\ket{-\vb w}$ to $\frac{1}{\sqrt{2}}\left[\ket{\vb w}\otimes\ket{-\vb w}-\ket{-\vb w}\otimes\ket{\vb w}\right]$, from the equator, with the wave function
\begin{equation}
	\ket{\psi}=\sqrt{\frac{1}{2}(1+\sin\alpha)}\ket{\vb w}\ket{-\vb w}-\sqrt{\frac12(1-\sin\alpha)}\ket{-\vb w}\ket{\vb w}
\end{equation}
Then letting $\omega^{(2)}=\i\braket{\dd\vb w|\dd \vb w}$, the Berry curvature of each site is $F_A^{(2)}=-F_B^{(2)}=\sin\alpha\,\omega^{(2)}$, and the flow of Berry curvature is $F^{(3)}_{AB}=-\cos\alpha\dd\alpha\,\omega^{(2)}$, exactly like the higher Berry curvature for the exactly solvable model. Several remark are necessary however. Firstly, this interpolation does not define a closed family of states, and supposing the Hilbert spaces have finite dimension, any interpolation to the south pole must have opposite net flow. Thus for finite system there are no new invariants, although the flow of Berry curvature is well defined, we need the boundaries at infinity to act as 'reservoirs' of the Berry curvature in order to find a nontrivial invariant. Secondly, for finite systems, it is not clear what constitutes a local variation. For two sites, there is no reason we shouldn't let $X_A$ entirely specify the wave function, and let $X_B$ be trivial. In this case there is no flow. The two site example is maximally ambiguous on this point, since for larger system, one could likely use some sort of correlation length argument to define a reasonable local parameter. It is interesting to note that we may construct a $d+1$ site model for the higher Berry flow in dimensions $d>1$, using the ideas of \cite{2024long}. Consider for example the $d=2$ three site $A,B,C$ case, with wave function $\ket{\psi}=\mc{H}_A\otimes\mc{H}_B\otimes\mc{H}_C$. Then there is a Schmidt decomposition between $\mc{H}_C$ and $\mc{H}_A\otimes \mc{H}_B$, $\ket{\psi}=\sum_\beta c_\beta\ket{\beta}_{[AB]}\ket{\beta}_{[C]}$, associating $X_C=\{c_\beta,\ket{\beta}_{[C]}\}$. For each $\beta$, let the Schmidt decomposition between
$\mc{H}_{A}$ and $\mc{H}_B$ be $\ket{\beta}=\sum_{\alpha} c_{\alpha;\beta}\ket{\alpha;\beta}_{[A]}\ket{\alpha;\beta}_{[B]}$. Then we define $X_A=\{\ket{\alpha;\beta}_{[A]}\},X_B=\{\ket{\alpha;\beta}_{[B]},c_{\alpha;\beta}\}$. Then the higher flow  $F_{ABC}^{(4)}=\dd_A\dd_B\dd_C\mc{A}$ is
\begin{equation}
	F^{(4)}_{ABC}=\i\sum_{\beta\alpha}\dd c_\beta^2\dd c_{\alpha;\beta}^2\braket{\dd\alpha;\beta|\dd\alpha;\beta}_{[A]}.
\end{equation}
This expression has the same issues as the two site expression for the $d=1$ flow for finite dimensional Hilbert spaces, but it is worth seeking out a regularization of the wave function analogous to MPS where this expression can be computed in infinite systems, since it and higher dimensional generalizations would give a natural definition of the higher Berry curvature in terms of the entanglement structure.

\let\i\oldi
\bibliography{MPSref.bib}
\end{document}